\documentclass[prl,aps,twocolumn,longbibliography]{revtex4-1} 
\usepackage{graphicx} 
\usepackage{color}
\usepackage{amsmath,amssymb,bm}
\usepackage{lipsum}
\usepackage{hyperref}

\usepackage{color}
\definecolor{bleuf}{rgb}{0,0.44,0.72}
\definecolor{bleuc}{rgb}{0.965,0.965,0.957}

\usepackage{hyperref}
\usepackage{amsmath}
\usepackage{amssymb}
\usepackage{lipsum}
\usepackage{placeins}

\hypersetup{
    colorlinks=true,                          
    linkcolor=bleuf, 
    citecolor=bleuf, 
    urlcolor=bleuf  }

\begin{document} 
\title{Sounds  and hydrodynamics of polar active fluids}
\author{Delphine Geyer}
\affiliation{Univ Lyon, ENS de Lyon, Univ Claude Bernard Lyon 1, CNRS, Laboratoire de Physique, F-69342 Lyon, France}

\author{Alexandre Morin}
\affiliation{Univ Lyon, ENS de Lyon, Univ Claude Bernard Lyon 1, CNRS, Laboratoire de Physique, F-69342 Lyon, France}

\author{Denis Bartolo}
\affiliation{Univ Lyon, ENS de Lyon, Univ Claude Bernard Lyon 1, CNRS, Laboratoire de Physique, F-69342 Lyon, France} 
\date{\today}

\maketitle

\textbf{Spontaneously flowing liquids have been successfully engineered from a variety of biological and synthetic self-propelled units ~\cite{Bausch2010,Dogic2012,Goldstein2013,Dogic2015,Sano2017,Nieves2017,Dauchot2010,Bricard2013,Granick,Sood2014,Sano2015}. Together with their orientational order, wave propagation in such active fluids have remained a subject of intense theoretical studies ~\cite{Toner95,Toner98,Tu98,Ramaswamy,CristinaPRX,AntonNatPhys}.  However, the experimental observation of this phenomenon has remained elusive. Here, we establish and exploit the propagation of sound waves in colloidal active materials with broken rotational symmetry. We demonstrate that two mixed modes coupling density and velocity fluctuations propagate along all directions in colloidal-roller fluids. We then show how the six materials constants defining the linear hydrodynamics of these active liquids can be measured from their spontaneous fluctuation spectrum, while being out of reach of conventional rheological methods. This active-sound spectroscopy is not specific to synthetic active materials and could provide a quantitative hydrodynamic description of herds,
flocks and swarms from inspection of their large-scale fluctuations ~\cite{Cavagna_Review,Ginelli2015,Locust,Midges_Natphys}.
}

\begin{figure*}
\begin{center}
\includegraphics[width=\textwidth]{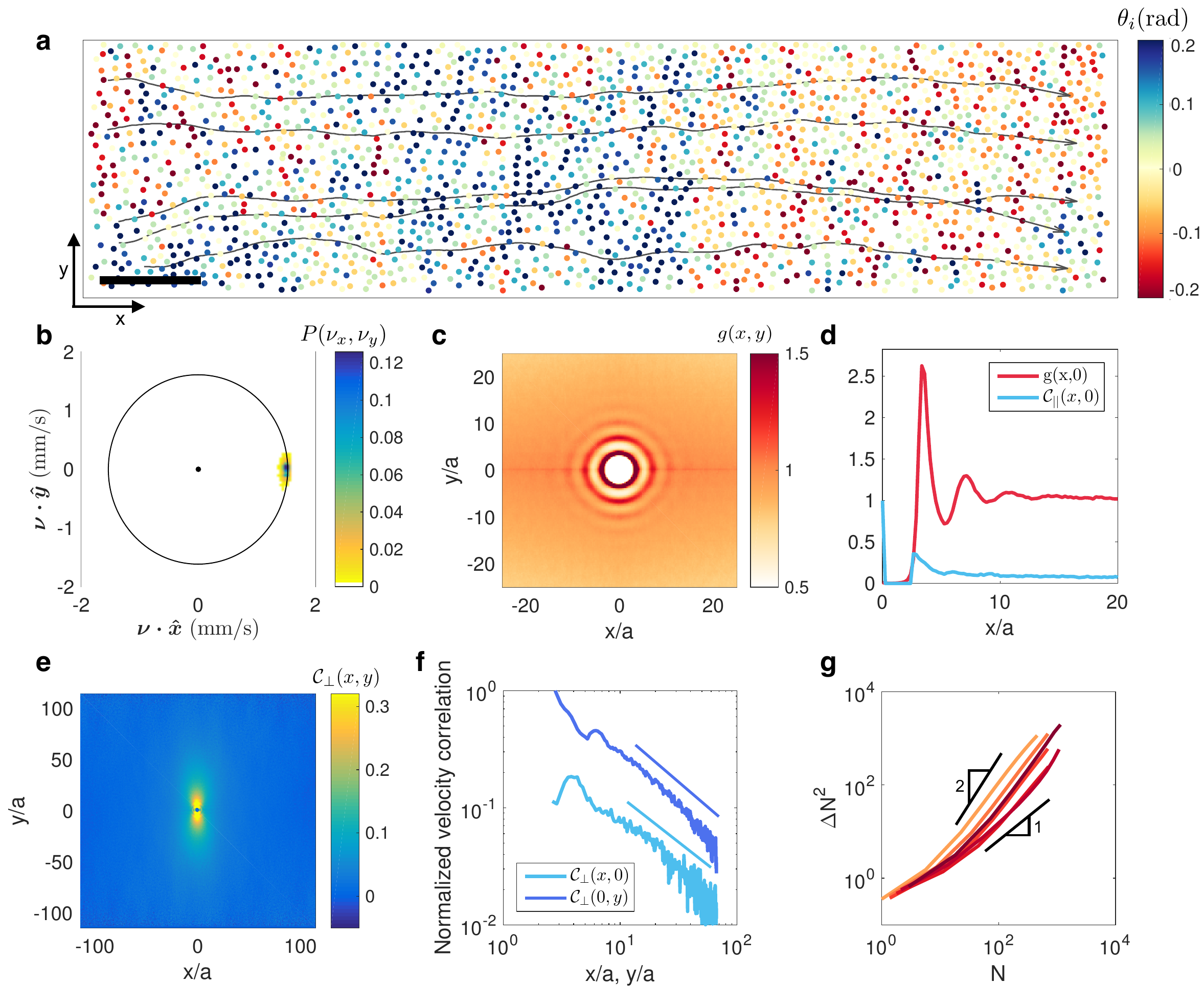}
{\color{black}
\caption{{\bf Colloidal rollers self-assemble into a spontaneously-flowing liquid.}
\textbf a, Close up on a microfluidic channel including $\sim3\times10^6$ colloidal rollers forming a homogeneous polar liquid. The color of the particles indicates the value of the angle, $\theta_i$, between their instantaneous velocity and the direction of the mean flow.  Five trajectories illustrate the typical motion of the rollers. $\rho_0=0.11$. Scale bar: $100\,\rm \mu m$. 
 \textbf b, Probability density function of the roller velocities, ${\bf \nu}_i(t)$, (ensemble and time integration). All the rollers propel along the same average direction.  $\rho_0=0.24$ as in all following panels.
\textbf c, The color indicates the value of the density pair correlation function $g(x,y)$ evaluated at  positions $(x,y)$. Structural correlations are short ranged and display only weak anisotropy.
\textbf d,  Cuts along the flow direction of  the pair distribution functions, $g(x,0)$~\cite{Hansen}, and of the longitudinal velocity correlations ${\mathcal C}_\parallel(x,0)$, where ${\mathcal C}_\parallel(\mathbf r)\equiv \langle \nu_i^\parallel(t)\nu_j^\parallel(t)\rangle_{( \mathbf r_i-\mathbf r_j)=\mathbf r,t}/\langle (\nu_i^{\parallel})^2(t)\rangle_{i,t}$. Both structural, and longitudinal- velocity correlations decay over  few particle radii.
 \textbf e, Correlations of the transverse velocity fluctuations (ensemble and time average): ${\mathcal C}_\perp(\mathbf r)\equiv \langle \nu_i^\perp(t)\nu_j^\perp(t)\rangle_{( \mathbf r_i-\mathbf r_j)=\mathbf r,t}/\langle (\nu_i^{\perp})^2(t)\rangle_{i,t}$. The transverse fluctuations are long ranged and strongly anisotropic.
\textbf f, The correlations of the transverse velocity fluctuations, ${\mathcal C}_\perp(\mathbf r)$,  decay algebraically in both directions. The solid lines correspond to best algebraic fits: ${\mathcal C}_\perp(x,0)\sim x^{-0.84}$, and ${\mathcal C}_\perp(0,y)\sim y^{-0.76}$.
\textbf g, Giant number fluctuations. Variance, $\Delta N^2(\ell)$, of the number of particles measured in  square regions of size $\ell$. $\Delta N^2(\ell)$ is plotted as a function of the average number of particle  $N(\ell)$ for five different polar active liquids of average area fractions $\rho_0=0.12,\,0.18,\, 0.18,\,0.24,\,0.30,\,0.39$ labeled by colors of increasing darkness. Solid lines:  scaling $\Delta N^2(\ell)\sim N(\ell)$ corresponding to normal density fluctuations as in equilibrium fluids, and $\Delta N^2(\ell)\sim N^2(\ell)$ scaling law predicted from linear hydrodynamic theory, see e.g.~\cite{Marchetti_Review}. Details about number fluctuation measurements and power-law fit values are provided in Supplementary Note 1.
}
\label{Fig1}
}
\end{center}
\end{figure*}

We exploit the so-called Quincke mechanism to motorize inert colloidal particles and turn them into self-propelled rollers~\cite{Quincke,Taylor69,Lavrentovich2016,Bricard2013}. 
When rolling on a solid surface they  interact via velocity-alignment interactions triggering a flocking transition as their area fraction exceeds $\rho_0=0.02$ ~\cite{Bricard2013}. 
As illustrated in Fig.~\ref{Fig1}a and Supplementary Video 1, millions of rollers interacting in a microfluidic channel self-organize to move coherently  along the same  direction, all propelling with the same average velocity $\langle {\bm \nu}_i\rangle= \nu_0\hat{\mathbf x}$, Figs.~\ref{Fig1}a and~\ref{Fig1}b. However, the flock  does not move as a  rigid body. Instead, it forms a homogeneous  active liquid  with strong orientational and little positional order,  Figs.~\ref{Fig1}b,~\ref{Fig1}c. 
Let us note $\nu_i^\parallel$ and $\nu_i^\perp$ the longitudinal and transverse components of the velocity fluctuations: $\bm \nu_i=(\nu_0+\nu_i^\parallel)\hat{\mathbf x}+\nu_i^\perp\hat{\mathbf y}$.
The correlations of the longitudinal  component, ${\mathcal C}_\parallel(\mathbf r)$, and of the liquid structure, are both short ranged and decay over few particle radii, Figs.~\ref{Fig1}c,~\ref{Fig1}d.  
In stark contrast, the correlations of the transverse velocity modes, ${\mathcal C}_\perp(\mathbf r)$, are anisotropic and decay algebraically, Figs.~\ref{Fig1}e and ~\ref{Fig1}f.  
This algebraic decay  
demonstrates that the transverse velocity fluctuations are soft modes associated with the  spontaneous symmetry breaking of the roller orientations~\cite{Toner_Review,Marchetti_Review}. In addition, self-propulsion couples these soft orientational modes to density fluctuations, thereby causing   the giant number fluctuations illustrated in Fig.~\ref{Fig1}g. Such anomalous fluctuations are  common to all orientationally ordered active fluids, see e.g.~\cite{Toner_Review,Marchetti_Review,Sano2017} and references therein. The density-fluctuation measurements,  and the discrepancy with~\cite{Bricard2013} are thoroughly discussed in Supplementary Note 1.

 Altogether these results establish that  colloidal rollers self-assemble into a prototypical polar active fluid. Their ability to support 
underdamped  sound modes, regardless whether the dynamics of their  microscopic units is overdamped, is one of the most remarkable, yet unconfirmed, theoretical prediction for  active fluids with broken rotational symmetry  ~\cite{Toner95,Tu98,Toner_Review,Marchetti_Review}. 
We  provide below an experimental demonstration of this counterintuitive prediction, and establish a generic method to measure the material constants of active fluids from their sound spectrum.

\begin{figure*}
\begin{center}
\includegraphics[width=0.9\textwidth]{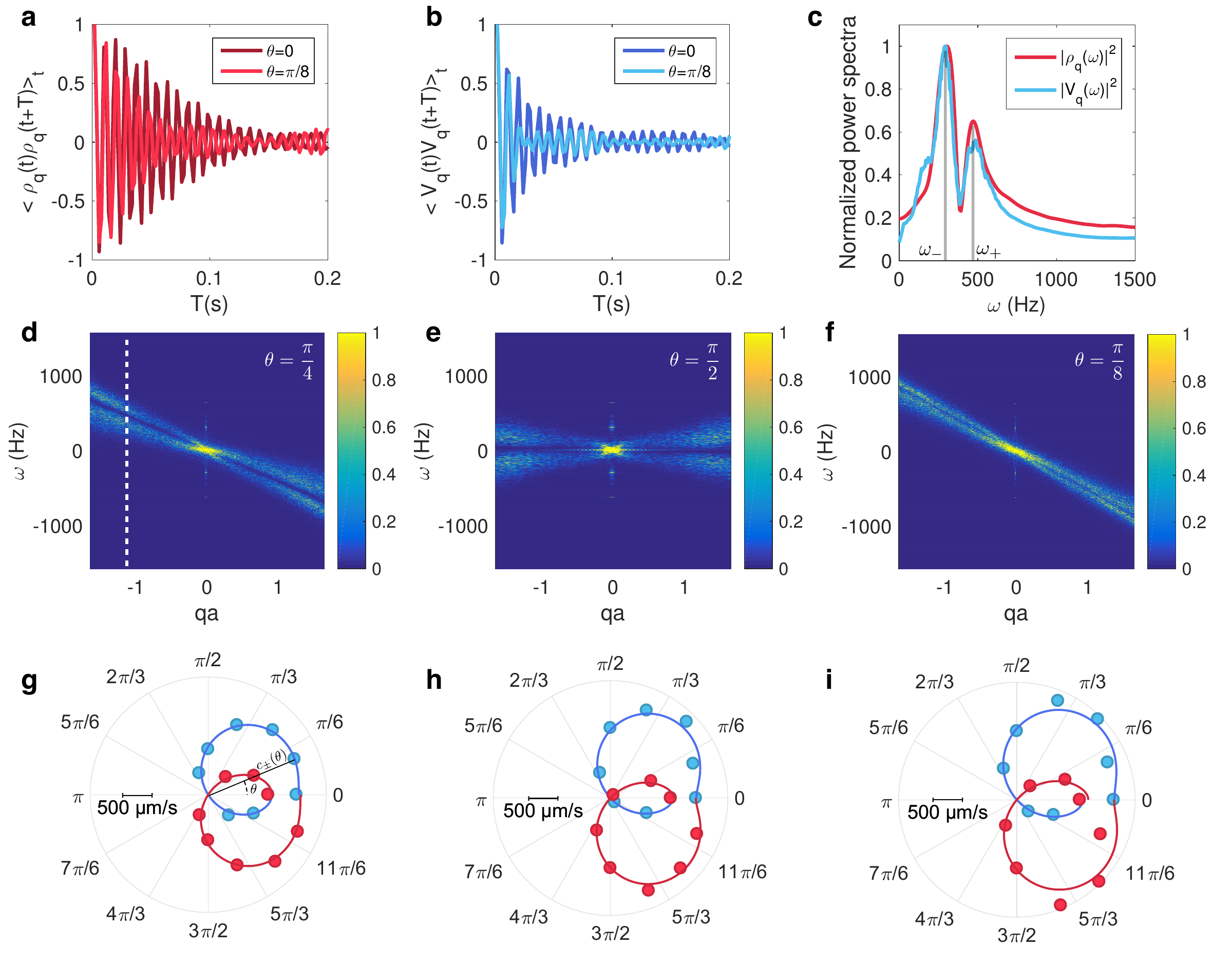}
\caption{{\bf Sound modes in polar active fluids.}
\textbf a, Two-time autocorrelations of the density fluctuations of wave vector $\mathbf q$, with $q=0.52\,\mu m^{-1}$, for two different directions of propagations, $\theta=0$ and $\theta=\pi/8$. $\rho_0=0.11$.
 \textbf b,  
 Two-time autocorrelations of the transverse velocity fluctuations for the same wave vectors as in $\mathbf a$. $\rho_0=0.11$.
\textbf c, Density (red) and velocity (blue) power spectra for $q=0.39$ and $\theta=\pi/4$. The two spectra have two peaks located at the same frequency $\omega_\pm$ and have identical width $\Delta \omega_\pm$. Both spectra reflect the propagation of the same mixed modes combining velocity and density excitations. $\rho_0=0.11$.
\textbf d, \textbf e and \textbf f, Full power spectra of the transverse velocity fluctuations $\langle|v_{q,\omega}|^2\rangle/\langle|v_{q=0,\omega=0}|^2\rangle$. They clearly show the dispersion relations of the mixed sound modes along three  different directions $\theta=\pi/4$, $\theta=\pi/2$ and $\theta=\pi/8$. The dashed line  in panel \textbf d corresponds to the cut showed in \textbf c. Sound modes propagate in a non dispersive fashion only at small $q$s. $\rho_0=0.11$.
\textbf g, \textbf h and \textbf i, Polar plots of the speed of sound, $c_\pm(\theta)={\lim_{q\to0}}[\omega\pm(\theta)/q]$ measured from the slope at $q=0$ of the dispersion relations. Experimental data: Red dots (resp. blue dots) correspond to  $c_+(\theta)$ (resp. $c_-(\theta)$). Solid lines: theoretical fits from Eq.~\eqref{Eq:c}. The roller area fractions are $\rho_0=0.11$ in \textbf g, $\rho_0=0.18$ in \textbf h, and $\rho_0=0.24$ in \textbf i.
\label{Fig2}
}
\end{center}
\end{figure*}

Let us consider the spatial Fourier components,  $\rho_\mathbf q(t)$ and  $v_{\mathbf q}(t)$, of the density and transverse velocity fields, where  $\mathbf q=q(\cos \theta,\sin\theta)$ is a wave vector making an angle $\theta$ with the mean flow direction, see Methods.  We show in Figs.~\ref{Fig2}a and~\ref{Fig2}b that the time correlations of $\rho_{\mathbf q}$ and $ v_{\mathbf q}$ oscillate over several periods before being damped, thereby demonstrating that both density and velocity waves propagate in the active liquid, see also Supplementary Videos 2 and 3. We emphasize, that we here consider the propagation of {\em linear} waves as opposed to the density fronts, or bands, seen at the onset of collective motion~\cite{Marchetti_Review,Bausch2010,Durian, Bricard2013}. 
In Fig.~\ref{Fig2}c the  power spectra  $|\rho_{q,\omega}|^2$ and $|v_{q,\omega}|^2$ evalutaed at $\theta=\pi/4$  both display two peaks located at identical oscillation frequencies  $\omega_\pm$. They define the frequencies of two mixed modes involving both density and velocity fluctuations intimately coupled  by the mass-conservation relation: $\partial_t \rho(\mathbf r,t)+\nabla\cdot\left[\rho(\mathbf r,t)\mathbf v(\mathbf r,t)\right]=0$. Repeating the same analysis for all wave lengths, we readily infer the dispersion relations $\omega=\omega_\pm(q)$ of the two sound modes as illustrated in Fig.~\ref{Fig2}d.   They both propagate in a dispersive fashion. Two speeds of sound can however be unambiguously defined at long wave lengths where: $\omega= c_{\pm}q$. Remarkably, both modes propagate in all directions and their  dispersion relation strongly depends on $\theta$, Figs.~\ref{Fig2}d, \ref{Fig2}e and \ref{Fig2}f. In particular, we find that the speed of sound $c_\pm (\theta) $ varies non-monotonically as shown in the polar plot of Fig.~\ref{Fig2}g. Measuring the angular variations of the speed-of-sound in active liquids with different area fractions, $\rho_0$, we find that the shape of the $c_\pm (\theta)$ curves is preserved, Figs.~\ref{Fig2}g,~\ref{Fig2}h,~\ref{Fig2}i and does not depend on the channel geometry, see Supplementary Note 3. Sound propagates faster in denser liquids.  

\begin{figure*}
\begin{center}
\includegraphics[width=\textwidth]{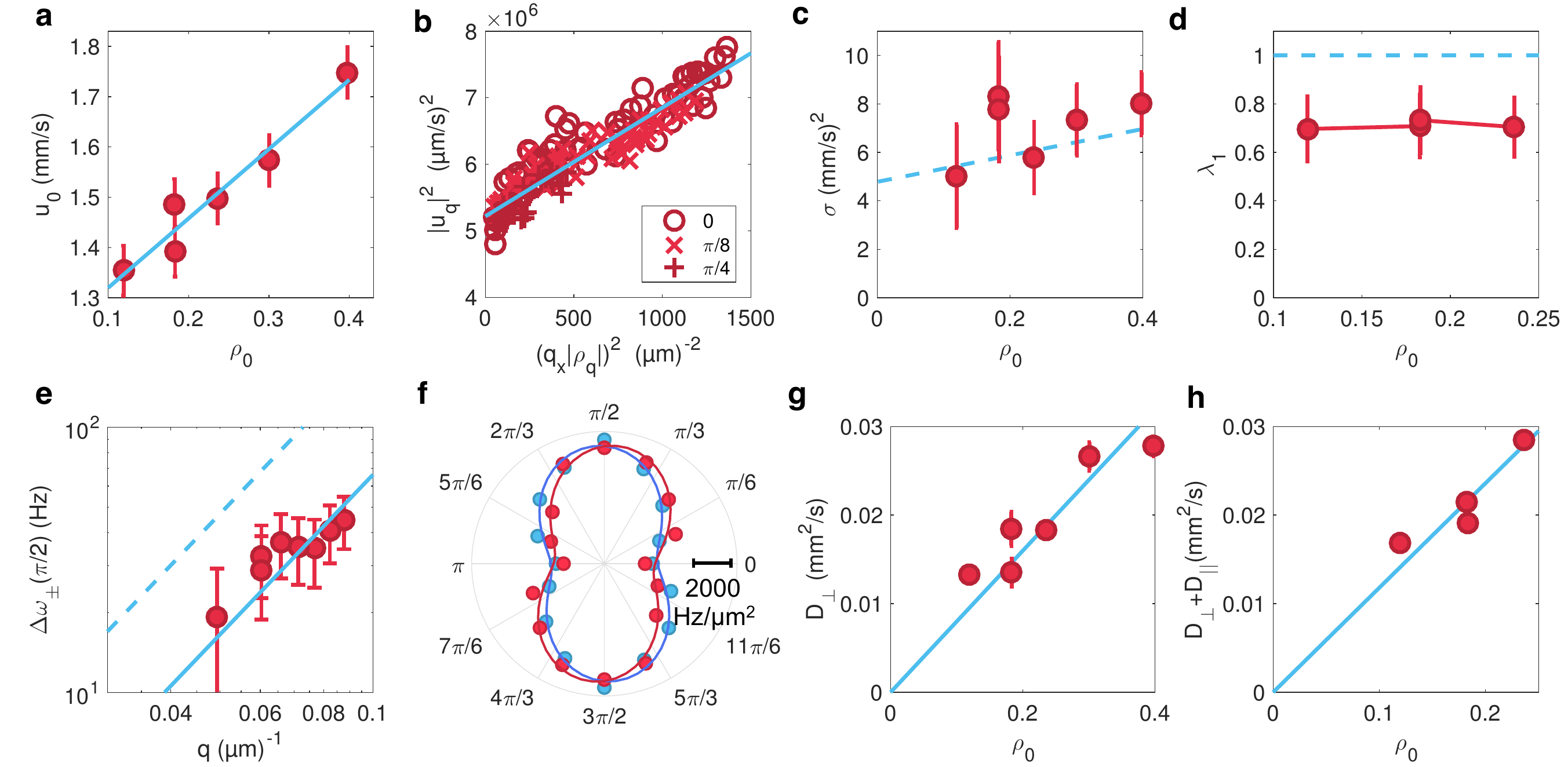}
\caption{{\bf Active-fluid spectroscopy.} Red dots: experimental data. Blue line: best linear fit. Dashed line theoretical prediction with no free fitting parameter deduced from kinetic theory, see Supplementary Note 2. The hydrodynamic description of the active fluid is inferred from:
\textbf a, Variations of the mean-flow speed with the mean area fraction.  Error bar: $100\,\rm\mu m/s$, 1std.  Denser fluids flow faster.
\textbf b, Parametric plot of the longitudinal velocity fluctuations $|u_q|^2$ varrying linearly with $(q_x|\rho_q|^2)$ for three propagation angles. The slope gives a measure of $D'=4\times10^{-6}\,\rm mm^2/s$. The offset at $q_x=0$ comes from the noise acting on the $u$ mode, see Supplementary Note 2. 
\textbf c,\textbf d,
The compressibility coefficient, $\sigma$ and advection coefficient, $\lambda_1$, are plotted versus the mean area fraction $\rho_0$. Both quantities are measured from the best fit of the speed of sound (Figs.~\ref{Fig2}g, ~\ref{Fig2}h, ~\ref{Fig2}).  The error bars defined applying the uncertainty-propagation formula on $\sigma=c_\pm({\pi/2})^2/\rho_0$ and $\lambda_1=c_+(0)/c_-(0)$. The uncertainties on $c$ and  $\rho_0$ are respectively $100\,\rm \mu m/s$ and $0.02$. 
\textbf e, Spectral width $\Delta \omega_\pm(\pi/2)$ of the modes propagating at $\theta=\pi/2$ plotted versus $q$ (log-log plot). $\Delta \omega_\pm(\pi/2)$ grows quadratically with $q$.  Error bars: $10\,\rm Hz$ estimated comparing several Lorentzian fits. Solid line: best quadratic fit. The bare prediction from the simplified kinetic theory overestimates $\Delta \omega_\pm(\pi/2)$ by a factor of 3. The possible origins of this overestimate are discussed in Supplementary Note 2. 
\textbf f, Polar plot of the spectral width normalized by $q^2$ and averaged over all wave vectors $\Delta_\pm=\langle\Delta\omega_\pm(\theta)/q^2\rangle_q$. Red (resp. Blue) dots: experimental data corresponding to $\Delta_+$ (resp. $\Delta_-$). Solid lines: best fits using the relation $\Delta_\pm=\frac{1}{4}[ -(D'+D_{\perp}+D_{\parallel})-(D'+D_{\parallel}-D_{\perp})\cos(2\theta)\pm D_{\rho}u_0\sqrt{{\rho_0}/{\sigma}} \sin(2\theta)]$, see Supplementary Note 2. 
\textbf g, Variations of the elastic constant $D_\perp$ with $\rho_0$. $D_\perp$ is measured from the quadratic fit shown in \textbf e, see main text.  Error bars defined as the $0.95$ confidence interval of the quadratic fit in \textbf e. The elastic constant increases linearly with the particle density.
\textbf h, Variations of the average elastic constant $D_\parallel+D_\perp$ with $\rho_0$. $D_\perp+D_\perp$ measured from the quadratic fit of $\Delta\omega_+(\pi/4)+\Delta\omega_-(\pi/4)$, see main text. Error bars defined as in \textbf g. 
\label{Fig3}
}
\end{center}
\end{figure*}

We now exploit these wave spectra to infer the hydrodynamics of the active-roller liquids from their spontaneous fluctuations.
The Navier-Stokes equations describing the flows of isotropic Newtonian liquids merely involve two material constants: density and viscosity. In  contrast,  in the absence of momentum conservation, and  given their intrinsic anisotropy, the hydrodynamics of polar active liquids  involve at least fourteen material constants, see e.~g.~\cite{Toner_Review,Toner2016} and Supplementary Note 2. We here focus on the linear dynamics of the density and velocity fields  around a homogeneous and steady flow along the $\hat {\mathbf x}$ direction.  We note $\rho(\mathbf r,t)$ the fluctuations of the density field around $\rho_0$, and  $\bm v(\mathbf r,t)=\left[u_0+u(\mathbf r,t)\right]\hat{\mathbf x}+v{(\mathbf r,t)}\hat y$ the local velocity field. As detailed in Supplementary Note 2, they evolve according to:
\begin{align}
&\partial_t{\rho}+\rho_0\partial_y v+u_0\partial_x \rho=D'\partial_x^2\rho,
\label{Eq:rho}\\
&\partial_t v + \lambda_1u_0\partial_x v=-\sigma \partial_y \rho+D_{\perp}\partial_{y}^2v+D_{\parallel}\partial_{x}^2v
\label{Eq:v}\nonumber\\
&+u_0D_{\rho}\partial_{xy}^2\rho,\\
&u=-\frac{D'}{\rho_0}\partial_x\rho.
\label{Eq:u}
\end{align}
Eq.~\eqref{Eq:rho}  corresponds to mass conservation, and Eq.~\eqref{Eq:v} describes the slow dynamics of the soft transverse-velocity mode. Eq.~\eqref{Eq:u} indicates that  $u(\mathbf r,t)$ is a fast mode. Longitudinal fluctuations quickly relax at all scales and are  slaved to $\partial_x\rho(\mathbf r,t)$, see  Supplementary Note 2 and~\cite{Toner95,Marchetti_Review,Baskaran2010}. The linear hydrodynamics of the active fluid is therefore fully prescribed by the emergent flow speed $u_0(\rho_{_0})$, Fig.~\ref{Fig3}a, and six material constants all having a clear physical meaning. $D'$ is a  diffusion constant readily measured from the linear relation between  longitudinal velocity fluctuations and density gradients defined by Eq.~\eqref{Eq:u} and confirmed by Fig~\ref{Fig3}b. $\lambda_1$ measures how fast velocity waves are convected by the mean flow and would be equal to one if momentum were conserved ~\cite{Toner98}. 
$\sigma$ is the active-liquid compressibility.  $D_{\perp}$ and $D_{\parallel}$  can either be thought as viscosities, or orientational elastic constants. Finally $D_{\rm \rho}$ stems for the couplings between  orientational and positional degrees of freedom between the active units.  Looking for plane-wave solutions of Eqs.~\eqref{Eq:rho} and ~\eqref{Eq:v}, we readily infer  the dispersion relations of mixed density and velocity waves. In the long-wave-length limit, they take the compact form predicted in~\cite{Toner98}: $\omega=\left[c_{\pm}(\theta)q+{\mathcal O (q^2)}\right]+i\Delta\omega_{\pm}(\theta)$, where $c_{\pm}$ is the speed of sound and the imaginary part $\Delta\omega_{\pm}(\theta)={\mathcal O}(q^2)$ corresponds to the  widths of the power spectra exemplified in Fig.~\ref{Fig2}c. The angular variations of the speed of sound are given by: 
\begin{align}
2c_{\pm}(\theta)&=\left(1+\lambda_1\right)u_0\cos\theta
\label{Eq:c}\\
&\pm\sqrt{\left(\lambda_1-1\right)^2u_0^2\cos^2\theta+4\sigma\rho_0\sin^2\theta}.\nonumber
\end{align}
This prediction is in excellent agreement with the  speed of sound measurements showed in Figs.~\ref{Fig2}g, \ref{Fig2}h, and~\ref{Fig2}i for three different densities.  As the mean-flow speed $u_0(\rho_0)$ is measured independently, fitting our data requires only two unknown functional parameters $\sigma(\rho_0)$ and $\lambda_1(\rho_0)$. The  variations of $c_\pm(\theta)$ therefore provide a direct measurement of the active-fluid compressibility and advection coefficients,~Figs.~\ref{Fig3}c and~\ref{Fig3}d.  The consistency of this method is further established by repeating the same measurements in two different channel geometries, and comparing  the density dependence of the hydrodynamic coefficients with the kinetic-theory predictions of ~\cite{Bricard2013,Bricard2015,Morin2017}, see Supplementary Note 2. Figure~\ref{Fig3}c shows a good agreement for the variations of $\sigma(\rho_0)$ over a range of densities. As in standard liquids, the compressibility increases with $\rho_0$. In the case of $\lambda_1$ the agreement is also satisfactory but not as accurate, see Fig.\ref{Fig3}d. Nonetheless theory  predicts the correct order of magnitude, and more importantly  the absence of variations of $\lambda_1$ with $\rho_0$. We now measure  the elastic constants of the active fluid from the damping of the sound waves.  Their damping time is set  by the inverse of the spectral widths $\Delta\omega_\pm=q^2\Delta_\pm(\theta)$, where the expression of the angular functions $\Delta_\pm(\theta)$ is given in Supplementary Note 2. Fig.~\ref{Fig3}e agrees with the $q^2$ scaling behavior, and we show in Fig.~\ref{Fig3}f that the angular variations of $\Delta_\pm(\theta)$ are correctly fitted by the linear hydrodynamic theory. Given the shape of the power spectra, Fig.~\ref{Fig2}f, measuring $\Delta_\pm$ at small $\theta$ is  out of reach of our experiments at high packing fractions. We therefore focus on two high angle values. 
 For $\theta=\pi/2$ and $\theta=\pi/4$, the spectral widths take the simple forms: $\Delta\omega_\pm(\pi/2)=q^2D_\perp/2 $  and  $\Delta\omega_\pm(\pi/4)=q^2\left[\frac{1}{4}(D_\perp+D_\parallel+D')\pm(\rho_0 D_\rho u_0)/\sqrt{4\sigma \rho_0}\right]$,
as detailed in Supplementary Note 2. A quadratic  fit of $\Delta\omega_\pm(\pi/2)$ therefore provides a direct measure of $D_\perp$, Fig.~\ref{Fig3}e. Similarly, a quadratic fit of  $[\Delta\omega_+(\pi/4)+\Delta\omega_-(\pi/4)]=\frac{1}{2}(D_\perp+D_\parallel+D')q^2$ gives the value of $(D_\perp+D_\parallel)$ as $D'=4\times 10^{-6}\,\rm mm^2/s$ is four orders of magnitude smaller than $D_\perp\sim D_\parallel\sim10^{-2}\,\rm mm^2/s$, see Fig~\ref{Fig3}b. 
The measured values of the elastic constants $D_\perp$ and $D_\parallel$  are shown in Figs.~\ref{Fig3}g and~\ref{Fig3}h for different packing fractions. Their order of magnitude, Fig.~\ref{Fig3}e, and more importantly their linear increase with $\rho_0$, Figs.~\ref{Fig3}g and~\ref{Fig3}h, are consistent with kinetic theory which also predicts that $D'$ should be vanishingly small.
 In principle, $D_\rho$ could be measured  for any polar active liquid  from the value of $\Delta\omega_+(\theta)-\Delta\omega_-(\theta)\sim D_\rho q^2 $. 
In the specific case of the colloidal rollers, kinetic theory predicts that $D_{\rho}$ should be independent of $\rho$. The precision of our measurements  is however not sufficient  for an accurate estimate of the variations of $\Delta \omega_\pm$ with the roller fraction.  For all fractions below $\rho_0=0.24$ we find $D_{\rho}=1\pm 0.5\,10\time10^{-2}\,{\rm mm}^2/s$. 
 Analysing the spontaneous fluctuations of the polar active fluids, we have  measured all its six materials constants, thereby providing a full description of its linear hydrodynamics. 

Before closing this letter,  two comments are in order. Firstly, the $q^2$ damping of the sound modes implies a $\Delta N^2\sim N^2$ scaling for the number fluctuations~\cite{Marchetti_Review}. While  giant number fluctuations are consistently found in all our experiments, linear theory overestimates their amplitude, see Fig.~\ref{Fig1}g.  This last observation might  suggest that the largest scales accessible in our experiments are smaller but not too far from the onset of  hydrodynamic breakdown predicted in ~\cite{Toner95,Toner98}.  Secondly, we  here focus  on homogeneous active materials. A natural extension to this work concerns sound propagation in more complex active media such as microfluidic lattices~\cite{AntonNatPhys}, or curved surfaces~\cite{CristinaPRX} where topologically-protected chiral sound modes are theoretically predicted.

In conclusion, two decades after the seminal predictions of Toner and Tu, we have experimentally demonstrated that the interplay between motility and soft orientational modes results in sound-wave propagation in colloidal active liquids. We have  exploited this counterintuitive phenomenon to lay out a generic spectroscopic method, which could  give access to the material constants of all active materials undergoing spontaneous flows. Active-sound spectroscopy  applies beyond synthetic active materials~\cite{SilentFlocks,MarchettiTurningFlocks}, and could be used to quantitatively describe large-scale  flocks, schools, and swarms as continuous media~\cite{Cavagna_Review,Ginelli2015,Locust,Midges_Natphys}.


\FloatBarrier
\newpage
\bibliography{bibliography_Delphine2017}

\noindent {\bf Acknowledgements.} We acknowledge support from ANR program MiTra and Institut Universitaire de France. We thank   O. Dauchot, A. Souslov and especially H. Chat\'e, B. Mahault, S. Ramaswamy, Y. Tu and J. Toner for invaluable comments and discussions. \\
\noindent{\bf Author Contributions.} D.~B. conceived the project.   D.~G. and D.~B. designed the experiments. D.~G. and A. M. performed the experiments.  D.~G. and D.~B. analyzed and discussed the results. D.~G. and D.~B. wrote the paper. \\
\noindent{\bf Author Information.} Correspondence and requests for materials
should be addressed to D.~B. (email: denis.bartolo@ens-lyon.fr).

\section*{Methods}
We use Polystyrene colloids of diameter $2a=4.8\,\rm \mu m$ dispersed in a 0.15 mol.L$^{-1}$ AOT-hexadecane solution (Thermo scientific G0500). The suspension is injected in a wide microfluidic channel made of two parallel  glass slides coated by a conducting layer of Indium Tin Oxyde (ITO) (Solems, ITOSOL30, thickness: 80 nm)~\cite{Bricard2013}. The two electrodes are assembled with double-sided scotch tape of homogeneous thickness ($H=110\,\mu \rm m$). More details about the design of the microfluidic device are provided in the  Supplementary Figure 5.

The colloids are confined in a $2\,\rm mm\times 30\, mm$ race track. The walls  are made of a positive photoresist resin (Microposit S1818, thickness: 2 $\mu$m). This geometry is achieved by means of conventional UV lithography.   After injection the colloids are let to sediment onto the positive electrode. 
Once a monolayer forms, Quincke electro-rotation is achieved by applying a homogeneous electric field transverse to the two electrodes $\mathbf E=E_0\hat{\mathbf z}$. The field is applied with a voltage amplifier (TREK 609E-6). All  reported results correspond to an electric field $E_0=2E_{\rm Q}$, where $E_{\rm Q}$ is the Quincke electro-rotation threshold $E_{\rm Q}=1\, \rm V/\mu m$.  The colloids are electrostatically repelled from the regions covered by the resin film thereby confining the active liquid in the racetrack. Upon applying $E_0$ the rollers propel instantly and quickly self-organize into a spontaneously flowing colloidal liquid. All  measurements are performed after the initial density heterogeneities have relaxed and a steady state is reached for all the observables. The waiting time is typically of the order of 10 minutes. 
 
 The colloids are observed with a 9.6x magnification with a fluorescent Nikon AZ100 microscope. The movies are recorded with a $4\,\rm Mpix$ CMOS camera (Basler ACE) at frame rates of 500 fps.  The particles are detected with a one pixel accuracy, and the particle trajectories and velocities are reconstructed using  the Crocker and Grier  algorithm [35].  Measurements are performed in a $1146\, \rm \mu m\times 286\,\rm \mu m$ observation window. The individual particle velocities are averaged over 4 subsequent frames.

The spatial Fourier transform of the density and transverse velocity fields are respectively defined as:
\begin{align}
\rho_{q}(t)&=\sum_i e^{iq \left[x_i(t)\cos \theta+y_i(t)\sin\theta\right]}, \\
\mathbf v_{q}(t)&=\sum_i {\bm\nu}_i(t)e^{iq \left[x_i(t)\cos \theta+y_i(t)\sin\theta\right]},
 \end{align}
 where  $x_i(t)$ and $y_i(t)$ are the instantaneous particle coordinates. The sum is performed over all  detected particles. Time Fourier transforms are then performed using the MATLAB implementation of the FFT algorithm.   The positions of the maxima, $\omega_\pm$, and the width $\Delta \omega_\pm$ of the velocity power spectra are determined by fitting the $|v_q(\omega)|^2$ curve by the sum of two Lorentzian functions. \\

\section*{Code avaibility } Matlab scripts used in this work are avaible from the corresponding author upon reasonnable request.

\section*{Data avaibility} The data that support the findings of this study are avaible upon request from the corresponding author.

\section*{Methods Reference}

[35] John Cocker and David Grier, "Methods of digital video microscopy for colloidal studies", Journal of colloid and interface science (1996).\\

\end{document}